\documentclass[twocolumn,showpacs,preprintnumbers,amsmath,amssymb]{revtex4}

\usepackage{epsfig}

\begin{document}
\title{An Iterative Procedure for the Estimation of Drift and
  Diffusion Coefficients of Langevin Processes} 
\author{D.~Kleinhans, R.~Friedrich}
\affiliation{Institute for Theoretical Physics, University of
  M\"unster, D-48149 M\"unster, Germany}

\author{A.~Nawroth, J.~Peinke}
\affiliation{ Institute for Physics, Carl-von-Ossietzky University Oldenburg, D-26111 Oldenburg, Germany} 
\date{\today}

\begin{abstract}
A general method is proposed which allows one 
to estimate drift and diffusion coefficients of a stochastic process
governed by a Langevin equation. It extends a previously devised
approach [R. Friedrich et al., 
Physics Letters {\bf A 271}, 217 (2000)], which 
requires sufficiently high sampling rates.
The analysis is based on an iterative procedure minimizing 
the Kullback-Leibler distance between measured and estimated 
two time joint probability distributions of the process.

\end{abstract}

\maketitle

\section{Introduction}
Complex behavior in systems far from equilibrium can quite often 
be traced back to rather simple laws due to the existence
of processes of selforganization \cite{Haken1}. 
Since complex systems are composed
of a huge number of subsystems, however, fluctuations stemming from
the microscopic degrees of freedom play an important role 
introducing a temporal variation on a fast time scale which quite
often can be considered as fluctuations. 
The consequence is the existence of evolution equations of a set of 
macroscopic order parameters ${\bf q}(t)$ which are governed by nonlinear
Langevin equations \cite{Risken}, \cite{Gardiner}:
\begin{equation}\label{Lange}
\frac{d}{dt}q_{i} = D_i^1({\bf q})  + \sum_l g_{il}({\bf q}) F_l(t)\quad ,
\end{equation}
where ${\bf q}(t)$ denotes the n-dimensional state vector, ${\bf
D}^1({\bf q})$ is the drift vector and the matrix $g({\bf q})$
is related to the diffusion matrix according to 
 $\left(D^2({\bf q})\right)_{ij}
=\sum_k g_{ik}({\bf q}) g_{jk}({\bf q})$. ${\bf F}(t)$ are fluctuating forces
with Gaussian statistics  
delta-correlated in time: $<F_l(t)>=0$, 
$<F_l(t) F_k(t')>=2\delta_{lk}\delta(t-t')$.
Here and in
the following we adopt It\^o's interpretation of stochastic 
integrals \cite{Risken}, \cite{Gardiner}. 

Analyzing complex systems, which can be described by
stochastic equations of the form (\ref{Lange}), therefore, amounts to
assess the underlying Langevin equations or the corresponding
Fokker-Planck equations from an inspection of experimentally
determined time series \cite{Haken2}. 
Recently, an operational method \cite{Siegert1}, \cite{Siegert2}
has been devised, which allows one to
estimate drift and diffusion coefficients of the 
stochastic processes from experimental data.
This method has been successfully applied to various problems in the
field of complex systems like the analysis of noisy electrical circuits
\cite{Siegert2}, stochastic dynamics of metal cutting 
\cite{Grad1}, systems with feedback delay \cite{Frank1},
meteorological processes like wind-driven Southern Ocean variability
\cite{Sura1}, traffic flow data \cite{Kriso} and physiological time series \cite{Kuusela04}. 
Furthermore it has been applied
to problems like turbulent flows \cite{PRL}, \cite{JFM}, 
passive scalar advection \cite{Tutku},
financial time series \cite{PRLfinanz}, analysis of rough surfaces
\cite{Jafari}, \cite{Waechter}, which can be characterized as a
stochastic process with respect to a scale variable exhibiting
markovian properties in scale.

The method is based on the evaluation of the time limits 
the first and second conditional moments, 
\begin{subequations}
\label{est}
\begin{eqnarray}
{\bf D}^1({\bf q}) &=& \lim_{\tau \rightarrow 0} \frac{1}{\tau}
< {\bf q}(t+\tau)-{\bf q}(t)|{\bf q}(t)={\bf q}> \\ 
{ D}^2_{ij}({\bf q}) &=& \lim_{\tau \rightarrow 0} \frac{1}{2\tau}
 < [{\bf q}(t+\tau)-{\bf q}(t)]_{i}\nonumber\\
&&[{\bf q}(t+\tau)-{\bf q}(t)]_{j}|{\bf q}(t)={\bf q}>\quad .
\end{eqnarray}
\end{subequations}
From these expressions it becomes evident that the sampling rate in the
experiments has to be sufficiently high in order to allow for a
reliable evaluation of the limit $\tau \rightarrow 0$.
Therefore, in all
applications mentioned above 
the results have been checked in a selfconsistent manner by a
recalculation of conditional pdf's from the estimated Fokker-Planck
equation. Possible problems in estimating drift and diffusion coefficients
related with low sampling frequencies have been adressed by Sura
\cite{Sura}, Ragwitz and Kantz \cite{Ragw}, \cite{Kantzcom} and 
Friedrich et al. \cite{Kantzrepl}.  

The aim of the present letter is to devise an extension of the above
method in order to overcome problems related with the time limit $\tau
\rightarrow 0$. These problems immediately show up for low
sampling rates.
 We also want to point out that for the case of stochastic forces
 ${\bf F}(t)$ with small but finite temporal correlations the process is not markovian in the
 limit $\tau \to 0$. In this case, however, one should use the Stratonovich
 interpretation of stochastic processes \cite{Risken}.



\section{Description of the Method}
The starting point is a first estimate of drift and
diffusion coefficients by the expressions (\ref{est}) evaluated for
the smallest reliably possible values of $\tau$. The second step
is an embedding of drift and diffusion coefficients into a family of
functions ${\bf D}^1({\bf q},\sigma)$, ${\bf D}^2({\bf q},\sigma)$
parameterized by a set of free parameters $\sigma$. The expressions
obtained in the first step 
already yield a crude estimate of the parameters $\sigma$. 
The third step consists in optimizing the free parameters
${\sigma}$. 

Optimization of the free parameters can be performed 
in the following way. One determines the 
conditional probability distribution
\begin{equation}
p({\bf q},t|{\bf q}_0,t_0;{\bf\sigma})
\end{equation}
for the parameter set ${\sigma}$ either by a
simulation of the Langevin equations or by a numerical
solution of the corresponding Fokker-Planck equation. In each case,
one can determine the two point pdf $f({\bf q},t;{\bf
q}_0,t_0;{\sigma})=p({\bf q},t|{\bf q}_0,t_0;{\sigma})f({\bf
q}_0,t_0)$ . 
The reader should note that this 
may be done for various finite values of $t-t_0$. The obtained two time
pdf can now be compared with the experimental one. A suitable measure
for the distance is the Kullback-Leibler information \cite{Haken2} 
 defined according to
\begin{eqnarray}
\label{kullb_information}
K({\sigma},t,t_0) &=&\int d{\bf q} \int d{\bf q}_0
f_{exp}({\bf q},t;{\bf q}_0,t_0)
\nonumber \\
&\times &
\ln \frac{f_{exp}({\bf q},t;{\bf q}_0,t_0)}{f({\bf q},t;{\bf q}_0,t_0,{\sigma)}}\qquad .
\end{eqnarray}

The minimum of the Kullback-Leibler information with respect to the parameters 
${\sigma}$ yields estimates of drift and diffusion of 
a stochastic process. This process is the best approximation
with respect to this measure in
the class of stochastic processes characterized by the parameters
${\sigma}$. The problem of identifying a stochastic process is then
equivalent to determining a minimum of the Kullback information. In practice
the minimum can be determined by gradient or genetic 
algorithms and solved by standard methods \cite{weinstein90}.
In the following we shall consider cases, where it
is possible to obtain a parametrization of the stochastic processes by 
only few parameters $\sigma$ such that the Kullback-Leibler measure
can be investigated by graphical means.

\section{Examples}
For certain classes of stochastic processes
the above procedure can be reduced considerably by the fact that
only few free parameters for the parametrization
of drift and diffusion terms have to be introduced. As a consequence 
the minimization procedure of the Kullback-Leibler information
is greatly facilitated.  

\subsection{One dimensional systems}

\begin{figure}
\begin{center}
\includegraphics[width=8.6cm]{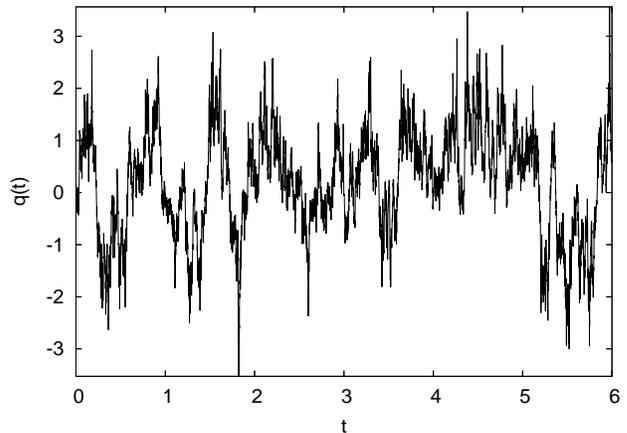}
\end{center}
\caption{Segment of the one-dimensional synthetic time series I. }
\label{mult01u2.dat}
\end{figure}

The case of one-dimensional systems allows for the following
treatment due to the fact that the 
stationary pdf, which is assumed to exist, can be determined
analytically:
\begin{equation}
f(q)=\frac{N}{D^{2}(q)} e^{\ \int\limits^q dq' \frac{D^{1}(q')}{D^{2}(q')}} \qquad .
\end{equation}
As a consequence, we have the relationship
\begin{equation}
\label{multnoise}
D^{1}(q)=D^{2}(q)\frac{d}{dq} \ln f(q)+\frac{d}{dq}D^{2}(q) \qquad . 
\end{equation}

Since $f(q)$ can be determined from the time series 
an estimate in terms of a parameterized ansatz 
for the diffusion term suffices. In fact, one may use the ansatz 
$D^2(q)=Q+ aq^2 +b q^4+\ldots$ , which 
helps in lowering the number of parameters $\sigma$ to be estimated by
the above procedure of minimization the Kullback-Leibler information.  
The drift then follows from (\ref{multnoise}). 

\begin{figure}
\begin{center}
\includegraphics[width=8.6cm]{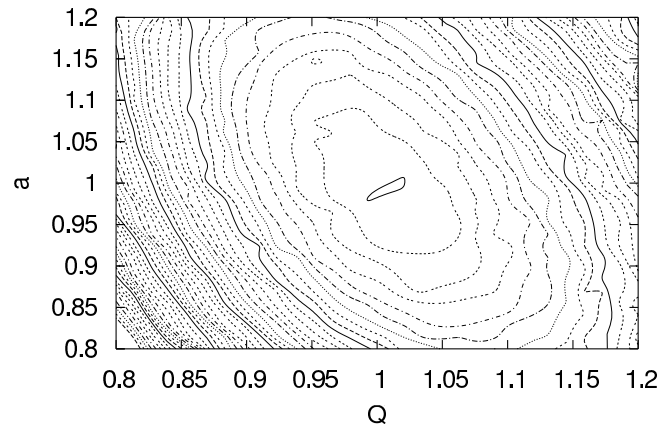}
\end{center}
\caption{Kullback distance $K(Q,a)$ as function of the parameters $Q$ 
and $a$ for time series I. The lines are equidistant 
contour lines starting from $2.6\cdot 10^{-4}$ in the center. 
The distance between contour lines is $5\cdot 10^{-5}$. 
A clear minimum is located at $(Q,a)=(1,1)$.}
\label{mult02.out}
\end{figure}

Let us consider system I with drift and diffusion functions
\begin{eqnarray}
D^1(q)=q-q^3\quad\mbox{and}\quad D^2(q)=1+q^2
\end{eqnarray}
driven by a multiplicative noise term.
We use synthetic data obtained by numerical integration of the 
corresponding Langevin equation \cite{Risken},
\begin{equation}
q(t+\tilde{\tau})=q(t)+\tilde{\tau}D^1\left[q(t)\right]+\sqrt{\tilde{\tau}}D^2\left[q(t)\right]\Gamma(t)\quad.
\end{equation}
A time series containing $10^6$ points with time 
increment $10^{-2}$ was generated. The intrinsic increment $\tilde{\tau}$ used for numerical integration 
of the corresponding Langevin equation was $10^{-5}$. 
A time segment of the data is presented in fig.~\ref{mult01u2.dat}. 
Since the stochastic process is stationary and ergodic  
all statistical quantities can be retrieved from this data.

For the estimation of the pdf's from data state space has to be 
divided into bins. We used $100$ equidistant 
bins for the stationary pdf. A very accurate way to calculate the 
integral yielding the Kullback-Leibler distance
without running out of memory even for higher dimensional 
data is to use an adequate local grid for the first argument (the
destination) of the conditional pdf's. The conditional pdf then 
locally can be retrieved from the data for any $({\bf q},{\bf q_{0}})$ 
with high accuracy. The local grid used in this  example covered $20$
equidistant bins.


During the iteration
procedure the two point pdf's have to be calculated. 
We again use the numerical simulation of  Langevin processes 
as a very  efficient way to generate these pdf's.

Starting from the estimates (\ref{est}) the ansatz $D^2(Q,a,q)=Q+aq^2$ 
is reasonable. The drift immediately follows from (\ref{multnoise}) 
and, for each parameter set $(Q,a)$,
one obtains a stationary distribution that equals the experimental one. 
Due to this fact the evaluation of the conditional 
pdf $p(q,t+\tau|q_{0},t;Q,a)$ suffices to calculate the Kullback-Leibler
distance. A clear minimum of the distance is found at $(Q,a)=(1,1)$ 
corresponding to the original set of parameters. 
The Kullback distance close to this minimum 
in the two-dimensional parameter space is 
exhibited in fig.~\ref{mult02.out}.

\subsection{Application to potential systems}

\begin{figure}
\begin{center}
\includegraphics[width=8.6cm]{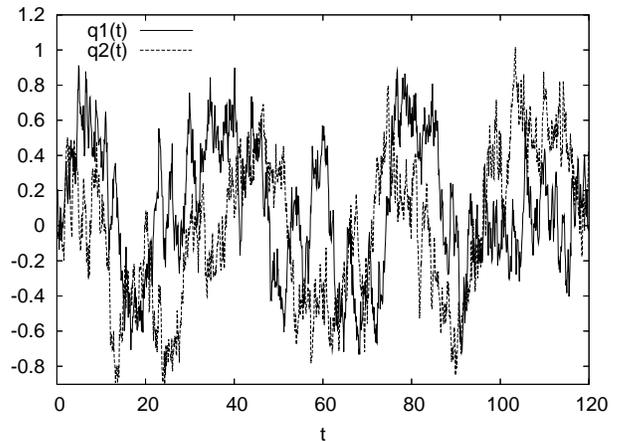}
\end{center}
\caption{Segment of the two-dimensional synthetic time series II.}
\label{feb002.dat.1d}
\end{figure}

\begin{figure}
\begin{center}
\includegraphics[width=8.6cm]{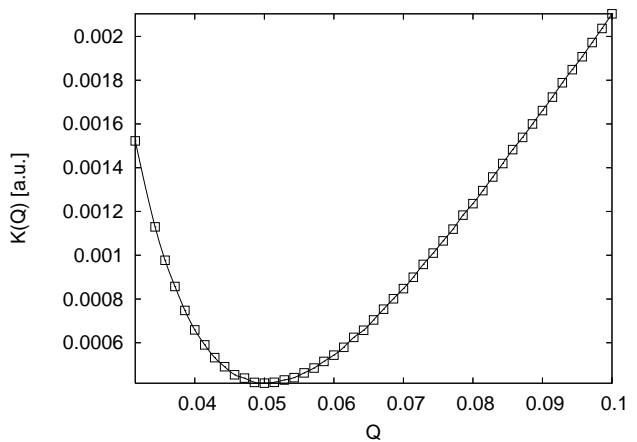}
\end{center}
\caption{The Kullback distance $K(Q)$ as a function of the 
noise strength $Q$ (time series II). A minimum is clearly visible at the value
$Q=0.05$.}
\label{feb002.out}
\end{figure}

The procedure for one-dimensional systems can be immediately 
extended to higher dimensions if one restricts the analysis
to the so-called class 
of potential systems for which the drift vector 
${\bf D}^1({\bf q})$ is obtained from a potential 
$V({\bf q})$ and $g_{ik}=\sqrt{Q}\delta_{ik}$.
The central point of our analysis is the following exact expression for
the stationary pdf 

\begin{equation}
f({\bf q})=N  e^{-V({\bf q}) /Q} 
\qquad .
\end{equation}

Since the stationary pdf can be estimated from experimental data 
one may parameterize the class of stochastic
processes by the single variable $Q$. Thus the drift function can be taken
to be fixed except for the value $Q$:

\begin{equation}
{\bf D}^{1}({\bf q})= Q {\bf \nabla} \ln f({\bf q}) \qquad . \label{add_final}
\end{equation}
 
As an example we consider the two-dimensional system
\begin{equation}
{\bf D^{1}}({\bf q})=
\left(\begin{array}{c}\epsilon q_{1}-q_{1}\left[q_{1}^2+Bq_{2}^2\right]\\
\epsilon q_{2}-q_{2}\left[Bq_{1}^2+q_{2}^2\right]\end{array}\right)
\qquad .
\end{equation}
This dynamical system arises as order parameter equations for instabilities
in nonequilibrium systems and has applications
from the fields of pattern formation in nonequilibrium systems to pattern
recognition \cite{Haken1}. It exhibits the features of 
multistability and selection. We considered the case
$\epsilon=0.25$ and $B=2$ (time series II). 
These parameters yield four stable fixpoints of the dynamics 
on the axes at $|{\bf q}|=1/2$ and unstable fixpoints at the 
origin and on the bisectional lines  at $|{\bf q}|=\sqrt{6}/6$. 

Data with time increments $10^{-1}$ 
for the datapoints 
has been generated with a time step 
$10^{-5}$ for the integration of the Langevin equations. 
The simulated time series II with $Q=.05$ consists 
of $5\cdot 10^{6}$ data points. 
Figure~\ref{feb002.dat.1d} exhibits a segment of the generated data.

We analyzed the time series as outlined above. 
State space in this case is divided in $100\times 100$ equidistant bins. 
Since the drift ${\bf D}^{1}({\bf q})$ can be evaluated 
from (\ref{add_final}) all parameters are fixed except for the 
noise strength $Q$.

After evaluating the Kullback measure for various values of 
$Q$ this value has to be optimized. The optimal value is
determined by the minimum of the 
Kullback distance. For the present case the minimum can easily be 
determined by graphical means.

\begin{figure}
\begin{center}
\includegraphics[width=8.6cm]{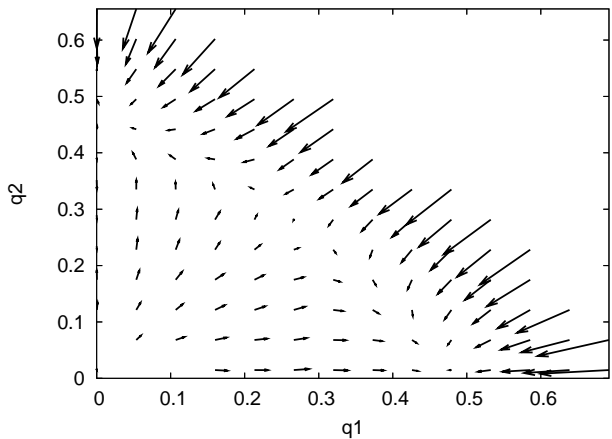}
\end{center}
\caption{Time series II: Drift vector field extracted from data
using the optimal value of $Q$. Unstable fixpoints in the center and 
on the bisectional line as well as 
the attractive fixpoints are clearly visible.}
\label{feb002.drift}
\end{figure}

Fig.~\ref{feb002.out} shows the Kullback distance
$K(Q)$ as a function of the noise strength Q for the time series II.
The minimum is clearly visible at $Q=0.05$ and agrees with the one 
used for simulation. With this parameter the drift 
vector field can be recalculated 
from the stationary distribution 
based on relation 
 (\ref{add_final}). 
The resulting drift vector field of dataset II is exhibited 
in fig.~\ref{feb002.drift}.

\section{Conclusion}

Summarizing, we have outlined an operational method for the estimation
of drift and diffusion terms from experimental time series of
stochastic Langevin processes. In contrast to previous approaches the 
present algorithm does not rely on estimating conditional moments in 
the small time increment limit. Although this limit 
yields a first
approximation an iterative refinement of the estimated stochastic
process is performed by minimization of the Kullback-Leibler distance between
estimated and measured two time probability distributions. 
The proposed procedure solves the problem of estimating drift and
diffusion terms of Langevin processes from time series. 
It involves the numerical solution of Langevin equations with
parameter dependent drift and diffusion terms, an evaluation of 
the Kullback-Leibler integral (which may be
determined by means of a Monte-Carlo method) and an 
optimization procedure, for which standard approaches
can be used. All involved steps are based on routine calculations.
Furthermore, restriction to certain classes of
stochastic processes like potential systems can drastically lower
the numerical efforts of the procedure. Therefore, the proposed
algorithm can be applied also to systems with higher dimensional
state spaces.

\end{document}